# Micro-Doppler-Coded Drone Identification

Dmytro Vovchuk[1,*], Mykola Khobzei[2], Vladyslav Tkach[2], Oleg Eliiashiv[3], Omer Tzidki[1], Konstantin Grotov[1], Aviel Glam[4] and Pavel Ginzburg[1]

[1]School of Electrical Engineering, Tel Aviv University, Tel Aviv, Israel
[2]Depertment of Radio Engineering and Information Security, Yuriy Fedkovych Chernivtsi National University, Chernivtsi, Ukraine
[3]Department of Information and Telecommunication Technology and Systems, Ivano-Frankivsk National Technical University of Oil and Gas, Ivano-Frankivsk, Ukraine
[4]Rafael Advanced Defense Systems Ltd., Haifa, Israel

**Abstract**

The forthcoming era of massive drone delivery deployment in urban environments raises a need to develop reliable control and monitoring systems. While active solutions, i.e., wireless sharing of a real-time location between air traffic participants and control units, are of use, developing additional security layers is appealing. Among various surveillance systems, radars offer distinct advantages by operating effectively in harsh weather conditions and providing high-resolution reliable detection over extended ranges. However, contrary to traditional airborne targets, small drones and copters pose a significant problem for radar systems due to their relatively small radar cross-sections. Here, we propose an efficient approach to label drones by attaching passive resonant scatterers to their rotor blades. While blades themselves generate micro-Doppler rotor-specific signatures, those are typically hard to capture at large distances owing to small signal-to-noise ratios in radar echoes. Furthermore, drones from the same vendor are indistinguishable by their micro-Doppler signatures. Here we demonstrate that equipping the blades with multiple resonant scatterers not only extends the drone detection range but also assigns it a unique micro-Doppler encoded identifier. By extrapolating the results of our laboratory and outdoor experiments to real high-grade radar surveillance systems, we estimate that the clear-sky identification range for a small drone is approximately 3-5 kilometers, whereas it would be barely detectable at 1000 meters if not labeled. This performance places the proposed passive system on par with its active counterparts, offering the clear benefits of reliability and resistance to jamming.

*corresponding author

# 1. Introduction

Drones and unmanned aerial vehicles (UAVs) become a resource for a wide range of civil applications, including, but not limited to infrastructure monitoring, airborne remote sensing, logistics, rescue operations, and many others [1]–[11]. Advanced navigation systems, high level of automatization, low cost, extended autonomous operation, and many other advantages promote extensive use of UAVs in the very near future and their impact will keep growing dramatically [12]–[19]. Nowadays, most flights conducted by UAVs occur in uncontrolled or segregated airspaces to eliminate danger. Nevertheless, small UAVs pose significant security issues, as has been proven in many unfortunate cases worldwide and the number of safety issues will continue to grow. Due to their low cost and unlicensed accessibility, small drones can be used by unauthorized users to carry dangerous items, spot on classified sites, interfere with air traffic, and for other undesired purposes. Apart from clear homeland security applications, it is evident that small drone traffic in an urban environment will grow exponentially in the near future, raising a clear need to deploy reliable monitoring systems [20]–[22].

Among the range of existent monitoring systems, the main approaches focus on camera, acoustic, and radar detection techniques, each having its own pros and cons [23]–[26]. While high-resolution imaging can allow for highly accurate target recognition, it heavily relies on a line of sight, ambient illumination, and intensive signal processing to name the key constraints. Acoustic approaches are rather range-limited and susceptible to environmental noises [27], [28]. Radar technologies have already proven themselves to operate in harsh conditions, providing real-time reliable detection of airborne targets [29]–[31]. This is the reason why those systems keep developing and are actively deployed on many different platforms, including automotive.

Small UAVs are a rather new class of airborne targets, which started to challenge radar technologies quite recently. Drones have enormously small radar cross-sections (RCS), hardly distinguishable from clutter. For example, birds with $10cm^2$ RCS at the X-band generate comparable radar signatures. In this case, an additional target classification must be performed. Micro-Doppler analysis, differentiating signatures of flapping wings and rotating blades comes at a rescue, though demanding a more accurate target investigation (e.g., more time on target and higher signal-to-noise ratio (SNR) in detection) [32]–[35]. Furthermore, another critical aspect is a low flight altitude, which puts clutter-filtering aspects at the cornerstone, as ground reflection and multi-path interference start playing key roles. State-of-the-art radar systems can detect small drones (e.g., DJI Mavic2) from kilometer-scale ranges in clear sky conditions and classify them with their micro-Doppler information from even shorter distances [36]. Considering civil applications, high-grade systems with KW-scale peak radiation power cannot be deployed in urban environments owing to a possible safety threat thus motivating to develop alternative approaches.

An effective way to enhance the radar visibility of small drones is to increase their RCS, such as by equipping them with corner reflectors. However, being also aerodynamically questionable, this approach is still insufficient in the case of heavy clutter. For example, a building behind a drone will always reflect more signal compared to the reflector. Considering the meter-scale range resolution (e.g., a typical 100MHz bandwidth at X-band) and proximity between targets, straightforward RCS enhancement does not provide a reliable solution. Furthermore, this approach does not grant a drone-specific identification number. On the contrary, given that spectral target-specific information is created, efficient clutter filtering can be obtained. The most common approach of separating a moving target from a static clutter is applying a Doppler filter or a moving target indicator, e.g., [37]. In both cases, mechanical motion is the key property, granting the capability to perform a long-range detection. Here, we propose to harness the inherent mechanical motion of the drone blades to implement an encoded transponder.

Apart from a conventional Doppler shift, a flying drone imprints special spectral features on radar returns. Those features are called micro-Dopler signatures of rotating blades. In general, micro-Doppler analysis is a technique used to extract additional information from radar returns by examining the Doppler shifts of different parts or components of a target's motion [38]–[40]. It is particularly useful for detecting and characterizing the movements of individual features or subcomponents within a larger target and is used across many disciplines [41]–[47]. Micro-Doppler analysis is heavily used nowadays for target classification [48]–[54]. However, in the case of small drones, classification, relying on micro-Doppler signatures of blades, required a significant time on target and relatively high SNR in detection. Our objective here is to get use of the fast-moving mechanical parts and dramatically enhance their micro-Doppler signatures by turning weakly scattering plastic blades into resonant scatterers, matched to an investigating radar bandwidth. For this purpose, we develop lightweight low-profile stickers and attach them directly to blades. As will be shown hereinafter, combinations of stickers at different quantities and places allow for coding information about a drone.

The manuscript is organized as follows: electromagnetic design and characterization of stickers come first and then followed by a set of indoor and outdoor experiments, to reveal the impact of stickers on the micro-Doppler signatures. Admitting the complexity of the observed signals, learning algorithms for target classification are applied, demonstrating long-range tagging capabilities. Extrapolation of the range, from which targets can be efficiently classified, is made with radar classification and probabilistic assessment.

## 2. Electromagnetic characterization of stickers

The design of the stickers is made to cooperate with an S-band radar, operating at 2-4GHz with a typical 100-200MHz bandwidth [36]. This band, being a compromise between the resolution and susceptibility to weather conditions, is a typical choice for airborne target surveillance and monitoring. As proof of concept, DJI Mavic2 drone was chosen for labelling. Worth noting that this item is a very difficult target for detection owing to its inherently small RCS.

The first set of studies concentrates on in-door analysis, performed in an anechoic chamber (Fig. 1(a)). The plastic blade of the drone is only 55mm long and, thus, can barely accommodate a resonant half-wavelength dipole (5cm at 3GHz) without meandering. Figure 1(b) demonstrates the photograph of the labelled drone, where Figs. 1(c) and (d) elaborate on the details. The meandered structure was designed to resonate at 3.35GHz (central frequency within the radar bandwidth). CST Microwave Studio was used for the modelling and simulations. Since the meandered dipole is a resonant structure, it is susceptible to the nearby dielectrics, specifically blades. To capture those effects, the optimization was done including the plastic blade (Fig. 1(c)). The dimensions and material features of the blade were taken from [55]–[57] - the permittivity $\varepsilon$ = 2.8, and dielectric loss $\tan(\delta)$ = 0.0054. Following the design, the stickers were manufactured by cutting the structure from a thin sticky foil sheet using an LPKF ProtoMat E33 milling and drilling machine. The RCS of the blade with the sticker was measured in an anechoic chamber. The results appear in Fig 1(e) and are in perfect correspondence to the modelling, demonstrating the pronounced peak at 3.35GHz.

At the next stage, the blade was placed back on the motor shaft of the drone. To recap on the operation principle, the micro-Doppler signature is the result of the time-dependent polarization mismatch between the dipole and the incident field polarization, which leads to the amplitude modulation of the scattered signal. Figures 1(g) and (h) demonstrate the concept, where the blade with the sticker is oriented differently with respect to the incident field polarization. The maximal scattering is observed in Fig. 1(g), while it is minimized in the case of (h). To prove this statement experimentally, the RCS of the entire drone, labelled with a single sticker, was acquired for 2 orthogonal orientations. The relevant value in this case is the differential RCS (the difference in scattering, which comes from the blade's orientation). The differential RCS spectrum appears in Fig. 1(f) and demonstrates a strong peak at exactly the same frequency, where the sticker has the resonance. This observation demonstrates that (i) the entire drone's body has a very small impact on the dipole response and (ii) the differential RCS value is $10^{-3}m^2$, which is comparable to the RCS of the entire structure. It suggests that if the target's Doppler is visible from more than 5 km, it can be classified with the aid of the sticker-controlled micro-Doppler from the same distance. This very strong statement, even though it comes from approximations, motivates further development of the proposed technology.

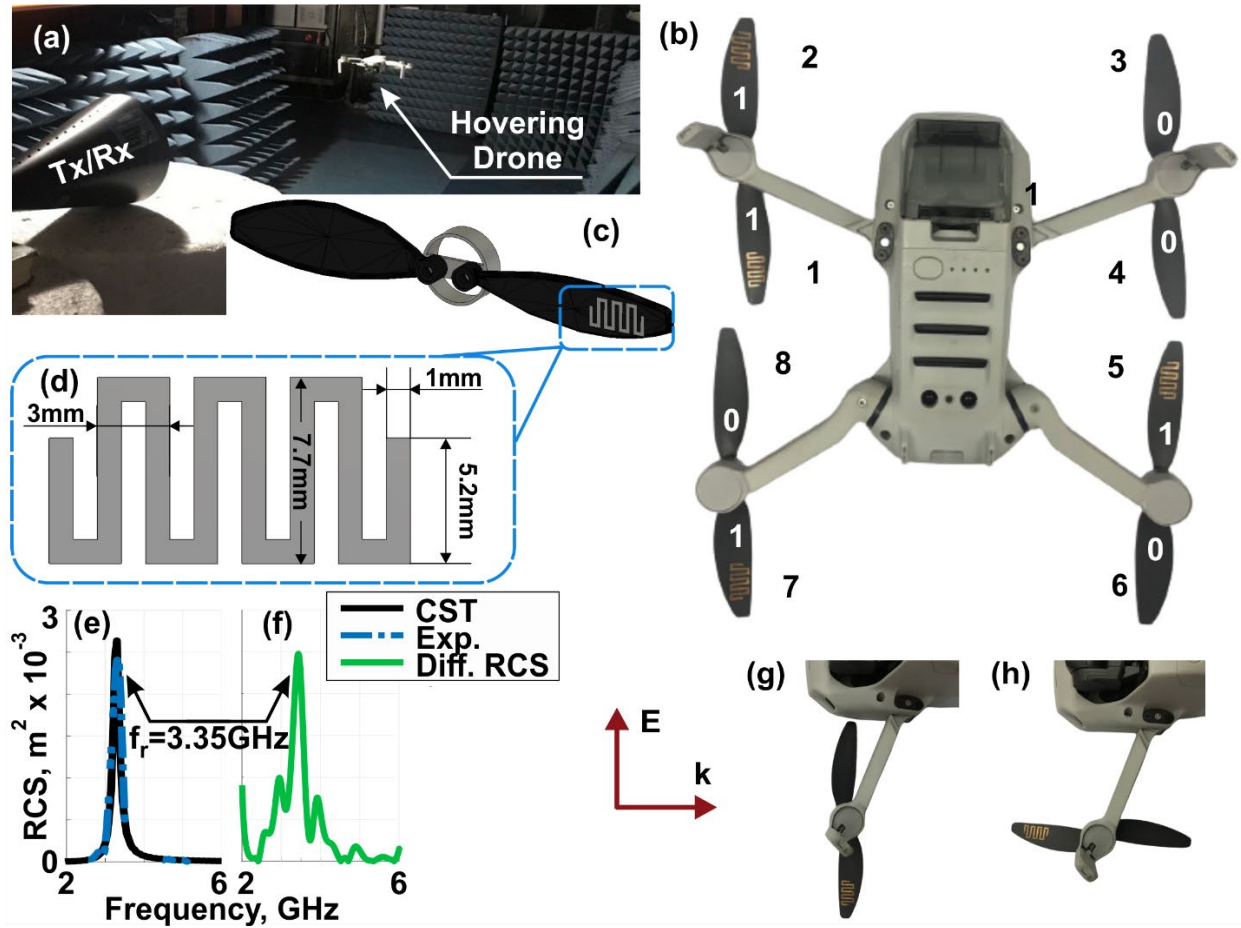

**Figure 1**. (a) Experimental setup - hovering drone in an anechoic chamber. CW radar is realized with a PNA and Tx/Rx antenna. (b) Photograph of the labelled drone, demonstrating the coding principle. Each blade has a serial number from 1 to 8, which is the basis for an 8-bit binary number. Each bit is either 0 (no sticker on the blade) or 1 (if the sticker is present). The code '11001010' is shown as an example. (c) CST layout of the meandered dipole on a blade. (d) Geometry of the optimized meander. (e) RCS spectrum of a sticker on the blade – experiment (blue dashed curve) and the numerical fit (black solid curve). (f) Differential RCS of the whole drone, with a single labelled blade, as appears in (g) and (h). Differential RCS is measured as a difference between the RCSs at (g) position subtracting (h) position of the rotor.

## 3. Micro-Doppler Coding

The setup, photographed in Fig. 1(a), was exploited to assess the micro-Doppler labelling capabilities and learn different sticker configurations (codes) at high SNR conditions prior to conducting outdoor experiments. The experimental setup consists of the PNA-based continuous wave (CW) radar and the hovering drone in front of the horizontally polarized radar's antenna, positioned at a 2-meter distance. The radar has a superior Doppler resolution, which is obtained by processing time traces of complex $S_{11}$ parameters, acquired by the PNA. The measurement was fully controlled with a PC, making the experimental data reproducible. Each experiment contained 10 seconds on target and 50001 measurement points within this time frame were acquired. Each point is the complex-valued reflection coefficient ($S_{11}$, peak-to-peak voltage) within a 100KHz filter around the carrier frequency. Considering the average angular velocity of the blades, the difference between adjacent measurements in time corresponds to 10 degrees of the blade turn. Those parameters were found sufficient for performing the subsequent signal processing.

As the first set of assessments, '10000000' and '11000000' codes were assessed. Figures 2 (a) and (b) demonstrate the spectrograms with an overlap between adjoining segments of 50% and a time window of 0.33 s, while Fig. 2(c) is the baseband spectrum of the radar echo, post-processed for the entire 10-second observation time. For the one-blade labelled case ('10000000' code), the baseband spectrum demonstrates a micro-Doppler comb, with equidistant ~160 Hz steps, which corresponds to the angular velocity of the rotor. The same behavior is seen in the spectrogram alongside a minor time evolution of the micro-Doppler spectrum. This behavior is expected as the drone hovers at the same place, nevertheless fluctuating around the stable point. When two blades of the same rotor are labeled ('11000000' code), the odd multiples of 160Hz become less visible (almost vanish) in the baseband spectrum. This is a rather expected result, as the blade becomes symmetric (has a radial symmetry with respect to the center of rotation). Thus, even harmonics predominate. This example provides an intuition on the coding mechanisms, which link the sticker configuration to the baseband micro-Doppler spectrum. For the sake of comparison, Figs. 2 (d, e, and f) demonstrate the same scenario for the outdoor experiment. While the outdoor scheme will be discussed in detail hereinafter, here it can be clearly seen that the spectrogram structure remains the same given the lines become wider. This phenomenon is solely related to the hovering stability. Outdoors, the drone has to contend with wind gusts that affect the rotor velocity, causing additional fluctuations to maintain the drone's position and rotational velocities.

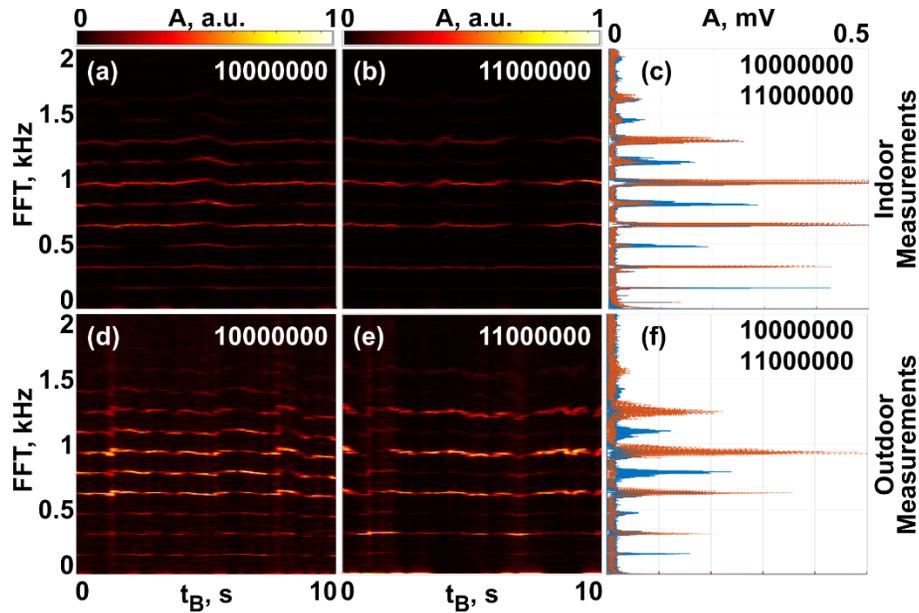

**Figure 2**. Micro-Doppler baseband spectra for a drone with one and two stickers on the same rotor. The corresponding codes are '10000000' and '11000000'. Indoor experiments – (a), (b), and (c). Outdoor experiments - (d), (e), and (f). Spectrograms – an overlap between adjoining segments of 50% and a time window of 0.33 sec. (c) and (f) are baseband micro-Doppler combs, obtained from 10-second time-on-target observation.

To reveal the significant differences between the codes, many other configurations were investigated. Several examples appear in Fig. 3, demonstrating both spectrograms and the micro-Doppler combs. Data acquisition was made under the same conditions as previously. The main result here is the complexity of the obtained signals, which even though can be visually distinguished from each other. However, the simple distinction between odd and even harmonics is not possible anymore. This complexity is the basis for machine learning algorithms, which will be investigated hereinafter on pathways to grant capabilities of remote labeling.

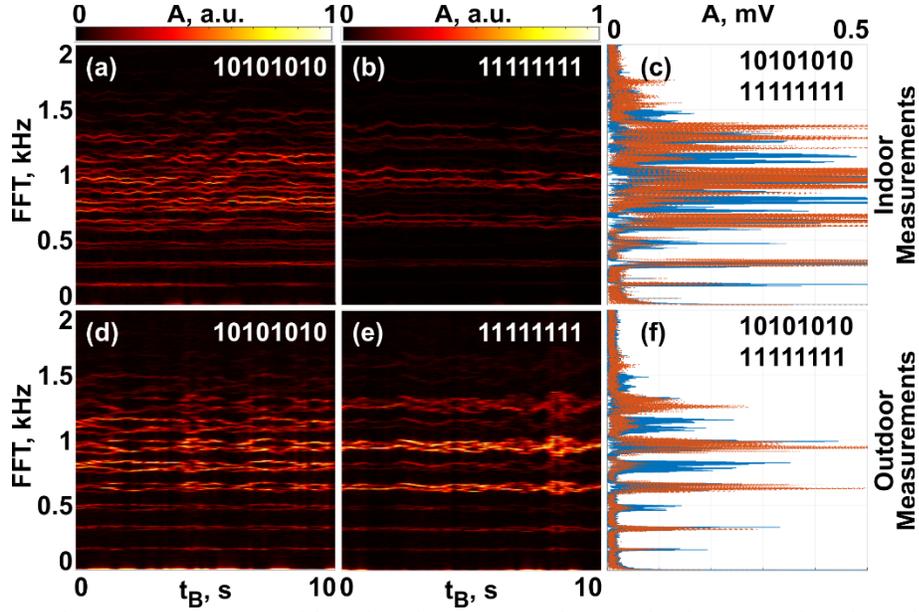

**Figure 3**. Micro-Doppler spectrograms and baseband micro-Dopler combs for several codes – indoor and outdoor scenarios. Details are in panels. Settings are the same as for Fig. 2.

## 4. Outdoor assessment of a moving drone

While hovering drone scenario allows obtaining high-SNR micro-Doppler signals and is very convenient for basic assessments, real flight conditions are quite different from this arrangement. To assess the impact of motion, albeit in a simplified arrangement, the drone was flown towards the radar antenna outdoors. Figure 4(a) is the photograph of the experiment, where '10000000' was used. Figure 4(b) demonstrates the spectrogram, obtained by postposing the experimental signal. 10 dBm CW signal was used. At t = 0 the drone started to fly towards the radar. The ground velocity was 1.3 m/s and the initial distance to the antenna was 10 m. Five main regions on the spectrogram can be identified – (I) the drone is too far and SNR is insufficient to make the detection, (II) SNR is sufficient to detect the moving target (via Doppler effect), but SNR is too low to see the micro-Doppler, (III) both Doppler and micro-Doppler are seen, (IV) the drone is hovering next to the radar antenna (0.2 m apart) and is still heading inward, and (V) the drone is hovering next to the antenna (1.5 meters apart), but it is heading outward.

Several important conclusions can be drawn from analyzing the data. The first one is on the perspective detectability range, which can be deduced from considering the radar equation. The range scales with the fourth root of the radiated power (Eq. 1) [58], [59]. Comparing the ~5 m detection with 10 dBm signal at our outdoor experiment and projecting the results on perspective high-grade radar, generating kW- MW peak signals, the proposed concept can enable classifying small drones at about $R_{max}$ = 3-5 km distance.

$$R_{max} = \sqrt[4]{\frac{P_s G^2 \lambda^2 \sigma}{P_{e_{min}}(4\pi)^3 L_{ges}}} \quad (1)$$

where $P_s$ – transmitted power [W]; G – Tx/Rx antenna gain (if the same antenna is used for transmission and receive) [linear scale]; λ central wavelength [m], c speed of light; σ – RCS [m²]; $P_{emin}$ – minimum receiving power [W]; $L_{ges}$ – loss factor [linear scale] that includes atmospheric, filter matching, transmit/receive and a few other possible loss mechanisms [60]. Parameters of real radar systems (e.g.,[36]) can be substituted into Eq. 1 and suggest even better detectability ranges. Specifically, the gain of our antenna is 10dB, while high-grade radars can have front ends of 30-40 dB.

The second observation is that for moderately slow-flying targets Doppler and micro-Doppler frequencies at the baseband are well separated. For example, kHz-scale frequencies correspond to 100m/sec velocity, which is very unlikely to emerge from a clutter. For example, the fastest birds, e.g., the peregrine falcon, can approach those but are unfortunately very unlikely to appear in urban environments.

The third conclusion is that the signals depend on a mutual orientation between the drone and the antenna. On one hand, it complicates the learning procedure on pathways to the classification, but, on another, it might allow extracting additional information given a sufficient SNR. Those aspects, however, are outside the scope of this first investigation in the series.

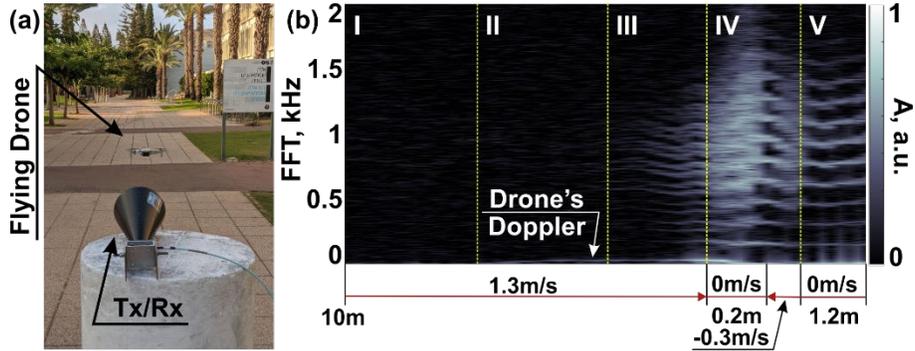

**Figure 4**. (a) Photograph of the outdoor experiment – low altitude flight in front of the radar antenna. (b) Spectrogram of the flying drone with '10000000' code. Different regimes are marked on the plot – (I) the drone is too far and SNR is insufficient to make the detection, (II) SNR is sufficient to detect the moving target (via Doppler effect), but SNR is too low to see the micro-Doppler, (III) both Dopler and micro-Doppler are seen, (IV) the drone is hovering next to the radar antenna (0.2 m apart) and is still heading inward, and (V) the drone is hovering next to the antenna (1.5 meters apart), but it is heading outward.

# 5. Target Classification and SNR Assessment

The capability to distinguish between the codes under different SNR conditions will be assessed next. Several classification approaches will be assessed and compared with each other on pathways to reveal a better methodology for perspective long-range radar measurements.

Quite a few methods to classify targets on their micro-Doppler signatures do exist. Time-frequency analysis is among the widely used ones. Typically, Short-Time Fourier Transform (STFT) or, more generally, Continuous Wavelet Transform (CWT)), enables following the evolution of Doppler frequencies over time [50]. These methods provide a spectrogram-like representation of the process and, in fact, STFT was used to produce Figs. 2, 3, and 4. Micro-Doppler analysis of swinging human arms and legs [52], [54], [61], helicopter blades, flapping bird's wings [62] was demonstrated by using this technique. The distinctive advantage of this methodology is the relationship between the physical model to the data, e.g., as was done in the discussion in Fig. 2. The drawback is the demand for a reasonable SNR, as will also become evident from the studies below. Machine Learning and Pattern Recognition techniques, such as neural networks, support vector machines (SVM), or clustering algorithms, can be employed to automatically detect and classify micro-Doppler signatures from raw data [48], [50], [61], [63]–[65]. The drawback of learning algorithms is the demand to acquire a large data set for the algorithm training and their relatively high susceptibility to a new type of data, for which the method was not trained. The latter issue opens room for opportunities for electronic warfare, for example. The choice of analysis technique depends on factors such as the specific application, the nature of the target and its movement, the available data, and the desired level of detail in the analysis.

Here we have chosen two methods to infer which is better for the target classification. For the proof-of-concept, four different codes (sticker arrangements – '00001111', '11111111', '10000000', '10101010'), will be distinguished by two classification algorithms. Laboratory measurements were considered to have a very high SNR, which is a mere approximation of reality, as the clutter is indeed suppressed, nevertheless, the receiver noise is still present. A part of the data set was used as a ground truth, while the rest was used for assessment. To assess the impact of noise on the detection probability, Gaussian white noise was intentionally added to the measurements. For making the statistical studies, 1000 numerical experiments with different noise realization have been performed. The correct classification probability was then extracted. Note that the search space here is very limited and contains only 4 classes. Consequently, the detection probability even at a very low SNR cannot fall below 25% of a completely random guess.

The first applied method is least squares analysis. The algorithm uses the FFT of the ground truth and compares it with noisy measurements using $\ell^2$ norm (integral over squared differences), applied to normalized signals. The classification is made by finding the shortest distance one between the four existing codes. The red curve in Fig. 5(b) demonstrates the result – while a very reliable detection is achieved at high SNR, the method stops working below 12dB noise level. It is rather expected from such a brute-force comparative method.

As another method for solving multi-label classification problems, we used a computer vision algorithm for analyzing spectrograms of the acquired raw data. The convolutional neural network (CNN) training was chosen, and preprocessing steps and augmentations were applied to the initial dataset. Firstly, 20 random subsamples of size 1000 were chosen from the initial time series, corresponding to 1 second of measurements. For each of the resulting time series, a spectrogram was generated and represented as a figure with a specific resolution of 500x500 pixels. As a result, a dataset of 460 annotated images was collected and used for training the neural networks. The convolutional neural network with a total of 12,845,828 parameters was designed. The Adam optimizer and the Cross-Entropy loss function were used. The model was trained on 80% of the laboratory data and tested on another 20%, where Gaussian white noise was added in the same way, it was done for the least squares analysis method. Spectrograms for four considered codes with different levels of additive Gaussian noise appear in Fig. 5(a). While Micro-Doppler frequencies can be clearly seen for high SNRs, they disappear in the noisy background. The blue curve in Fig. 5(b) demonstrates the detection probability. As expected, the graphs converge for random classification of 25% for low SNRs and for ultimate classification for high (above 20dB) SNRs. Furthermore, the CNN algorithm outperforms least squares analysis at lower SNR, which is quite important in the case of outdoor radar measurements. The algorithm can be further improved with hyperparameter tuning, increasing the number of network parameters and more data collection of the spectrograms, hence using fewer synthetic data.

To assess the applicability of the method in the real environment, we applied it to the outdoor data. Those measurements (Fig. 4(a)) are quite noisy and performed in a heavy clutter (low flying altitudes). Furthermore, 7 different codes (instead of 4) were assessed. The same data augmentation process analysis was performed and validated on 20% of the dataset, which wasn't involved during training. The method demonstrated 71.4% accuracy on the validation set, highlighting the need for intensive data acquisition, which is well-known in the field of target classification.

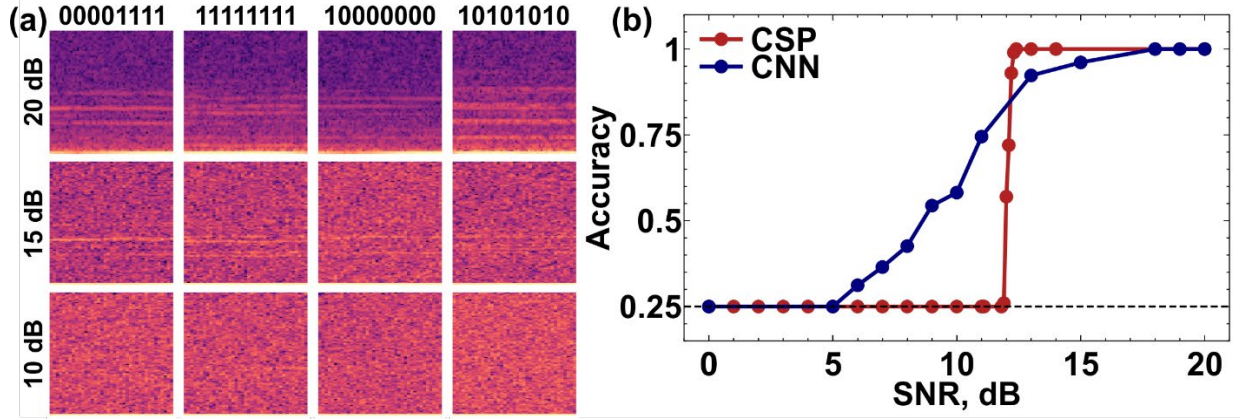

**Figure 5.** Micro-Doppler codes classification. (a) Micro-Doppler spectrograms for 4 different codes, obtained at different SNRs (synthetic data). (b) Detection probability. Red curve - least squares analysis, blue - convolutional neural network, applied on a set of images (examples are in panel (a)).

**Outlook and Conclusion**

A solution for the passive labeling of small UAVs has been proposed and demonstrated. The demand for extended reliability in ground control monitoring systems, which utilize radar-based sensors, necessitates the use of passive methods. Compared to active solutions that aim for continuous reporting of UAV coordinates, passive architectures do not overload wireless traffic and are less susceptible to jamming.

The approach is based on tagging the drone's blades with resonant stickers, which efficiently interact with electromagnetic radiation. Driven into rotary motion, the resonators generate geometry-specific micro-Doppler signatures, significantly elevating them compared to those generated by untagged samples. Those target-specific signals become detectable from large distances, which can approach 3-5km given high-grade radar monitoring systems are in use. This new approach was applied to a small drone, tagged with several distinguishable micro-Doppler codes, which were assessed in both laboratory and outdoor conditions, thus verifying the methodology. Considering the perspective of micro-Doppler tagging degrees of freedom, the proposed method can compete with active transponders and even provide solutions in case of heavy clutter, low-flying altitudes, and urban environments. These aspects will become critically important in future applications, where small drones are anticipated to play a pivotal role in delivery missions, transforming logistics and transportation dynamics.

**Acknowledgments:**
Israel Innovation, Authority, Department of the Navy, Office of Naval Research Global, under ONRG Award N62909–21–1–2038, Israel Science Foundation (ISF grant number 1115/23). M.K. acknowledges the support from IEEE Antennas and Propagation Society Fellowship Program (2023).